\documentclass[prl,twocolumn,aps,showpacs]{revtex4} % showkeys

\usepackage{graphicx}% Include figure files
\usepackage{dcolumn}% Align table columns on decimal point
\usepackage{bm}% bold math

\newcommand{\realset}{{\bf R}}
\newcommand{\realsetplus}{{\bf R}^{+}}
\newcommand{\glog}{\Lambda}
\newcommand{\gexp}{{\cal E}}

\newcommand{\mcol}{\quad,\quad}

\newcommand{\ep}{\epsilon}

\begin{document}

\title{Generalized Boltzmann factors and the maximum entropy principle}
%\shorttitle{Generalized Boltzmann factors}

\author{Rudolf Hanel$^{1,2}$}
\email{Rudolf.Hanel@ua.ac.be}
\author{Stefan Thurner$^{2}$}
\email{thurner@univie.ac.at}
\affiliation{
   $^1$Vision Lab; University of Antwerp;	
   Groenenborgerlaan 171; B-2020 Antwerp, Belgium\\
   $^2$Complex Systems Research Group  HNO; Medical University of Vienna; W\"ahringer G\"urtel 18-20; A-1090 Vienna; Austria \\   
}

%\pacs{05.90.+m}{Other topics in stat. phys., thermodynamics, and nonlin. dyn. systems}

\begin{abstract}
We generalize the usual exponential Boltzmann factor to any reasonable and potentially  
observable distribution function, $B(E)$. By defining generalized logarithms $\glog$ as 
inverses of these distribution functions, we are led to  a generalization   
of the classical Boltzmann-Gibbs entropy, 
$S_{BG}= -\int d\ep \,\, \omega(\ep) \,\, B(\ep) \,\, \log B(\ep)$ to  the expression 
$S\equiv -\int d\ep \,\, \omega(\ep) \int_0^{B(\ep)} dx \,\, \glog (x)$, which   
contains the classical entropy as a special case. 
We demonstrate that this entropy has two important features: 
First, it describes the correct thermodynamic relations of the system, 
and second, the observed distributions are straight forward solutions to the Jaynes
maximum entropy principle with the ordinary (not escort!) constraints.  
Tsallis entropy is recovered as a further special case.   
\end{abstract}

\maketitle

\section{Introduction}

It has been realized, that many statistical systems in 
nature can not be satisfactorially described by naive or 
straight forward application of Boltzmann-Gibbs statistical mechanics. 
In contrast to ergodic, separable, locally and weakly interacting systems,  
these systems are {\it complex} systems 
whose characteristic distributions often are of power-law type. 
Due to the existence of strong
correlations between its elements complex systems
often violate ergodicity and are 
prepared in states at the {\it edge of chaos}, i.e. they exhibit
weak sensitivity to initial conditions. 
Further, complex systems are mostly not separable
in the sense, that probabilities for finding a system in a given state 
factorize into single particle probabilities and as a consequence, renders 
these systems not treatable with Boltzmann single particle 
entropies \cite{ludwig}. 
However, it is evident that Gibbs entropies can in principle 
take into account any correlations in a given system, as the 
full Hamiltonian $H$, including potential terms,  
enters. Since in the following we will be only concerned about measurable 
quantities in statistical systems we will take the 
Gibbs entropy as a starting ground 
\begin{equation}
 S_G=
  - \int d\Gamma \,\, B \left(H(\Gamma)\right) \,\, 
  \log\left(B\left(H(\Gamma)\right)\right) \quad ,
\label{gibbs}
\end{equation} 
where $\Gamma$ are the phase space variables, and $B$ is 
the Boltzmann factor, which usually reads, 
$B(H)\sim \exp(-\beta H)$, for the canonical distribution.
It is interesting to note that the exponential 
form of the Boltzmann factor is not a priori dictated 
by classical statistical mechanics, but that much of   
classical statistical mechanics is built upon 
this special form of the Boltzmann factor, as argued e.g. in 
\cite{kaniadakislog}.

Classical statistical mechanics was designed 
for systems with short- (or zero-) range interactions, 
such as gas-dynamics. The exponential was found to be 
the natural choice in countless systems. However, for extending the concept 
of statistical mechanics to complex systems, which are characterized by 
fundamentally different distribution functions,  it seems natural to allow
generalizations of the Boltzmann factor. 
What is the Boltzmann factor?
What are the minimum requirements and restrictions 
to call some function a Boltzmann factor? 

The normalized Boltzmann factor is a probability to encounter 
a particular  state 
in the bath system, representing the hidden physical influences 
the observable 
ensemble of properties are subject to and thus
closely relates to experiment.
In the canonical ensemble the density of states with energy $E_1$ 
are given by 
\begin{equation}
 \rho(E_1) =\omega_1(E_1) \omega_2(E-E_1)Z^{-1} \quad ,
\end{equation}
where $\omega_1$ is the subjective microcanonical density,
i.e. the multiplicity of states in the ensemble of observable
properties, and $\omega_2$ is the bath density. 
$E$ is the energy of the total system, which is usually
unknown, and $Z$ is the partition function. 
Usually, the normalized $\omega_2(E-E_1)Z^{-1}$ is identified with 
the Boltzmann factor. However, in this form it explicitly depends on 
the total system energy $E$. This total energy should be factored out into 
a multiplicative factor since 
measured quantities should not depend on $E$. This factor 
will be canceled by $Z$, which is of course $E$ dependent. If the Boltzmann 
factor is taken as an exponential, this separation is trivial. 
Another approach is to ask which classes of Boltzmann factors 
allow for such a factorization. The answer was given in \cite{hanel1}, 
showing by a mathematical argument, that the most general Boltzmann factors which 
allow for an $E$ separation are of so-called $q$-exponential type.

In the following,  we start by  exploring a most general form of the 
Boltzmann factor, compatible with the requirements of normalizability, 
monotonicity and the possibility of $E$ separation. 
We do not fix the specific form of this factor
which (in principle) can be determined from measurements. 
We ask whether one can construct a theoretical framework
where data, i.e. the measured distribution serves as a starting point 
to construct an entropy which is consistent with both, the 
correct thermodynamic relations and the Jaynes maximum entropy principle \cite{jaynes}. 
According to this modification of logics it is 
sensible in a first step to modify or deform the $\log$ in Eq. (\ref{gibbs}) 
to a generalized logarithm $\Lambda$. 
The concept of deforming logarithms and thus modifying the 
form of entropy in order to accommodate 
a large body of experimental data from complex systems is not new
\cite{tsallis88,tsallis05,abelog,naudts,wada,kaniadakislog,borges,filho}.
An axiomatic definition of generalized logarithmic and exponential 
functions $\glog$ and $\gexp$ has been given
in \cite{naudts_physA2002} where also the concept of dual logarithms of the form 
$\glog^{*}(x)\equiv-\glog(1/x)$ has first been introduced. 
An algebraization of the deformed concept, i.e.
$x\otimes y=\gexp(\glog(x)+\glog(y))$, and $x\oplus y=\glog(\gexp(x)\gexp(y))$, has been given in
\cite{kaniadakislog}, where this structure has been exploited in the context of special relativistic
mechanics. In \cite{kaniadakis_physrefE2005} a constrained variational principle has then been utilized with respect
to trace-form entropies deriving a family of three-parameter deformed logarithms $\log_{(\kappa,r,\zeta)}$, being the 
most general of its kind so far, containing -- to our best knowledge -- all possible logarithms that are compatible
with the standard variational principle $\delta \tilde G=0$, with the usual functional
\begin{eqnarray}
\begin{array}{lll}
 \tilde G = \tilde S_G[B] &-& \beta \int d\ep \,\, \omega\left(\ep\right) B \left(\ep\right)(\ep-U) \\
  &-&\gamma \left( \int d\ep \,\, \omega\left(\ep\right)  B(\ep) - 1\right)   \quad ,
\end{array}
\label{stdvarprinciple}
\end{eqnarray} 
with the generalized Gibbs entropy 
\begin{equation}
\tilde S_G[B] = 
- \int d\ep \,\, \omega\left( \ep \right) B \left(\ep\right) \,\, \glog\left(B\left(\ep\right)\right) \quad ,
\label{gge}
\end{equation}
where $U$ is the measured average energy, $\omega(\ep)$ is the multiplicity,  
$\beta$ is the usual inverse temperature, and $\gamma$ is the Lagrange parameter
for normalizability.
%The logic in this approach is to start from the variational principle
%and explore the Boltzmann factors which can be used consistently
%with the variational principle. 
%In this sense it is possible to identify the constraint term of the 
%variation of $G$, with a Boltzmann factor of the 
%form $B=\bar \alpha \glog^{-1}(1/\lambda(-\beta(E-U)-\gamma-\bar \eta))$, i.e. 
%by equating $-\gamma-\beta(E-U)=\lambda\glog(B/\bar \alpha)+\bar \eta$. 
%$\lambda$, $\bar \alpha$, and  $\bar \eta$ are real constants related to the 
%parameters $\kappa,r,\zeta$ parametrizing the three-parameter 
%generalized Kaniadakis logarithm.
%This defines a differential equation with respect to $\glog$, \cite{kaniadakis_physrefE2005}. 
%The solutions of this equation are then considered candidate 
%generalized logarithms. 

The novel logics of this paper is that we  start from a measured distribution, the  
Boltzmann factor, which is not necessarily of standard exponential form. 
We want to keep the intuition of the origin of the Boltzmann factor as the adequately normalized
contributions of the bath, i.e. we require $\rho(E)=\omega(E)B(E)$, where $\omega$ 
is the multiplicity of the energy state in the observable system and represents 
our knowledge about the experimental device we observe in order to retrieve data. 
In principle, $\omega$ can be known which makes the Boltzmann factor $B(E)=\rho(E)/\omega(E)$ 
factor indirectly measurable. 
To keep close formal contact with usual statistical physics, we represent the measured Boltzmann 
factor by replacing the usual exponential function by some function $\gexp$, i.e.
\begin{equation}
  \exp(-\beta(E-U)-\tilde{\gamma}) \to \gexp(-\beta(E-U)-\tilde{\gamma}) \quad ,
\end{equation}
where $\tilde{\gamma}$ is the normalization constant. 
We then construct an entropy such that two requirements are strictly fulfilled: 
First, the entropy leads to the correct thermodynamics of the system, and 
second, the Jaynes variational principle holds.
%In the classical setting, $\tilde{\gamma}=\log(Z)$, 
%and we would thus also like to preserve the expression $S_G=\tilde{\gamma}$ in the generalized case.
%We show that this is indeed possible for a straight forward generalization of (\ref{gibbs}). 
%As we will see, it is possible to interpret $\tilde{\gamma}=\glog^*(Z)$ in the generalized 
%setting for a natural definition of the generalized partition function $Z$.
%Further, we show that a slight 
%modification to the Jaynes variational principle \cite{jaynes} is 
%necessary to also derive the expected relations from a variational 
%approach. 
%We conclude with two examples how -- for a given (measured) form of 
%a Boltzmann factor -- the deformed logarithm is determined. 

\section{The generalized Boltzmann factor}

Let us start by listing three ''axioms'' containing the 
intuitively clear minimum requirements for a 
Boltzmann factor $B$, 
\begin{enumerate}
\item $B$ is monotonous and positive. 
\item $B$ can be normalized, i.e. $\int d\ep \, \omega_1(\ep) \, B(\ep)=1$. 
\item $B$ must not explicitly depend on the total system energy. 
It must be possible that 
the $E$ term in the argument of 
$\omega_2(E-E_1)$ can be factored out, i.e., 
%\begin{equation}
$
\omega_2(E-E_1)= F(E-E^*) \,\, B(E_1-E^*) $, 
%\quad ,
%\label{sep}
%\end{equation}
where the normalized version of $B$ we shall call a Boltzmann 
factor. $F$ is some function, and $E^*$ some reference energy, 
e.g. the equilibrium energy. In \cite{hanel1} and \cite{hanel2} 
an explicit program was shown how this separation is 
uniquely obtained. 
\end{enumerate}

We thus write a Boltzmann factor which fulfills all requirements 
\begin{equation}
B(H) \equiv \gexp (-\beta(H-U)- \tilde{\gamma}) \quad , 
\label{boltz}
\end{equation}
where $\tilde{\gamma}$ is the normalization constant (partition function),
$U$ and $\beta$ being  the measured average energy 
and inverse temperature, respectively. 
Monotonicity 
and positivity are assumed to be properties of the
generalized exponential functions $\gexp$, which then 
implies the existence of inverse functions, 
the associated generalized
logarithms $\glog=\gexp^{-1}$. From a generalized 
logarithm $\glog$ and its dual ($\glog^{*}(x) \equiv -\glog(x^{-1})$) 
one assumes the usual properties, 
\begin{equation}
 \begin{array}{l}
  \glog:\realsetplus\to\realset  \\ 
  \glog(1)=0\mcol\glog'(1)=1 \mcol \glog'>0 \\ 
  \glog''<0 \quad \quad \mbox{(convexity)} \quad , 
  \end{array}
\end{equation}
implying analogous properties for the generalized exponential function.

%\section{The Generalized Gibbs Entropy}

Now, with any representative of the above allowed generalized Boltzmann factor $B$ and 
its associated logarithm $\glog$
let us in a first step generalize Gibbs entropy Eq. (\ref{gibbs}), (same  as Eq. (\ref{gge})),  
\begin{equation}
  \tilde S_{G}\equiv -\int d\Gamma B\left(H(\Gamma)\right)
  \glog\left(B\left(H(\Gamma)\right)\right) \quad ,
\label{entroptilde}
\end{equation}
and compute the Gibbs entropy as follows 
\begin{equation}
\begin{array}{ccl}
\tilde S_{G}&=&-\int d\Gamma \,\,
				B\left(H\right)\glog\left(B\left(H\right)\right)  \\				
     &=&
      - \int d\ep \int d\Gamma 
       \delta\left(\ep-H\right) 
				B\left(\ep\right)\glog\left(B\left(\ep\right)\right) \\
     &=&\int d\ep  \,\, \omega_{H}(\ep) \,\, 
        \gexp\left(-\beta (\ep-U)-\tilde{\gamma}\right) 
      \left(\beta (\ep-U)+\tilde{\gamma}\right),
\end{array}
\label{ge}
\end{equation}
where $\omega_{H}(E) \equiv \int d\Gamma \delta(E-H)$ is the microcanonic 
multiplicity factor for the energy $E$ which represents the observable system.
As a shorthand notation we will write Eq. (\ref{ge}) as in Eq. (\ref{gge}), 
$\tilde S_{G} = \int d\ep \,\, \omega(\ep) \,\, B(\ep) \,\, \glog(B(\ep) )$, 
with $B(E)=\gexp(-\beta(E-U)-\tilde \gamma)$. 
With the definition of the expectation value 
\begin{equation}
\left<f\right> \equiv 
\int d\ep \, f(\ep) \, \omega_{H}(\ep) \, \gexp\left(-\beta (\ep-U)-\tilde{\gamma}\right)\quad,
\end{equation}
it becomes obvious that the normalization constant $\tilde{\gamma}$ has to be chosen
such that 
\begin{equation}
  \int d\ep \,\, \omega_{H}(\ep) \,\,
  \gexp\left(-\beta (\ep-U)-\tilde{\gamma}\right) = 1\quad.
  \label{normalization_A}
\end{equation}
Using this and specifying $\left<\ep\right>=U$, we get 
  $ \tilde S_{G}=\tilde{\gamma}$.  We drop the subscript $H$ in the following. 
Looking at $\tilde S_G$ for $\beta=0$, implies that $B(E)=Z^{-1}=\mbox{const}$, for
$Z=\int d\ep \,\, \omega(\ep)$, and therefore
$\tilde S_G=-\int d\ep \,\, \omega(\ep)Z^{-1}\glog(Z^{-1})=-\glog(Z^{-1})$. 
Thus one identifies
\begin{equation}
 \tilde S_{G}= \tilde{\gamma}=-\glog(Z^{-1})=\glog^*(Z) \quad .
 \label{dual}
\end{equation}
 Note, that to get a finite $Z$ it is necessary to understand 
the integral $\int d\ep \omega(\ep)$, in the limits $E_1=0$, and $E_2=E_{\rm max}$, 
where $E_{\rm max}$ is the largest energy of the observable system.
Such regularizations are of course implicitly present under all experimental circumstances. 
If we wish this relation to hold for all $\beta$ it is interesting
to observe that the partition function $Z$ also has to be defined in a deformed way, i.e. 
using the definition of the deformed product 
$x\otimes y=\gexp(\glog(x)+\glog(y))$, analogous to \cite{kaniadakislog}.
The renormalization condition can then be recast into the form
\begin{equation}
 B(H)=
 \left(\frac{1}{Z}\right)\otimes
 \gexp\left(-\beta \left(H(\Gamma)-U\right)\right) \quad , 
 \label{boltz2}
\end{equation}
which becomes the defining equation for the generalized partition function $Z$. 
In this sense the generalization of Boltzmann factors 
naturally involves dual logarithms, whose occurence 
has been noted recently in the context of generalized entropies
\cite{wada,naudts,borges,tsallis05}.
This is of course 
just of relevance for non self-dual logs, examples of which include the 
$q$-logarithm ($\log_{q}^{*}(p)=\log_{2-q}(p)$) and the Abe-log \cite{abelog}.

\section{The variational principle}

Using the standard variational principle  Eq. (\ref{stdvarprinciple}) on the basis 
of the generalized entropy given in Eq. (\ref{ge}) (with the usual constraints!), 
the only possible choice for $\glog$ is the ordinary $\log$. To see this, variation of 
Eq. (\ref{stdvarprinciple}) yields  
\begin{equation}
\frac{d}{dB}  B \glog(B) = -\gamma-\beta(E-U) \quad .
\label{res}
\end{equation} 
By substituting $B=\gexp\left(-\beta(\ep-U)- \tilde \gamma) \right) $, it is clear that the only 
solution to this is $\glog(B) = \log(B)$, and $\gexp$ can thus only be 
the ordinary exponential Boltzmann factor.  
This is unsatisfactory. 
 
The problem arises because for any generalized $\glog$ other than the ordinary $\log$
there exists a non-trivial extra term, $B\glog'(B)$, in Eq. (\ref{res}).  In order to cancel 
this term we suggest to further generalize the generalized logarithm $\glog(B)$ to a 
{\it functional} in the following way, 
\begin{equation}
\glog(B) \rightarrow \bar \glog[B]\equiv   \glog(B) - \eta[B] \quad ,
\end{equation} 
where we use $[ B ]$ to indicate  functional dependence on $B$. 
By substituting  $\glog$ by $\bar \glog$ in Eq. (\ref{gge}), we obtain the entropy 
%
%To solve the problem we suggest the following modification of 
%the entropy $\tilde S_G$. Let us add a constant, $\eta[B]$, which is functionally dependent 
%on the measured distribution $B$, and define
\begin{equation}
S[B] \equiv  \tilde S_G [B] + \eta[B] \quad ,
\label{entr}
\end{equation}
where we have used that $\eta$ is a constant with respect to $\epsilon$-integration and 
the normalization condition (\ref{normalization_A}). 
Now the idea is that after  variation with respect to $B$, the additional term 
$\frac{\delta}{\delta B} \eta[B]$, can be used to cancel the term $-\omega(E) B(E) \frac{d}{dB} \glog(B(E))$. 
The corresponding condition, $\frac{\delta}{\delta B} \eta[B]=\omega(E) B(E) \frac{d}{dB} \glog(B(E))$, 
implies the form of $\eta$ 
%where $\eta[B]$ is chosen such that the Jaynes maximum entropy principle works. 
%We use $[ B ]$ to indicate  functional dependence on $B$. 
%As an Ansatz we take
\begin{equation}
\eta[B] = \int d\ep \,\, \omega(\ep) \int_0^{B(\ep)} dx \,\, \glog'(x) x   + c \quad ,
\label{ansatz}
\end{equation}
Let us substitute this into Eq. (\ref{entr}) to get
\begin{equation}
 \begin{array}{lll}
  S[B] &=& \eta[B]   - \int d\ep  \,\, \omega(\ep) \,\, B(\ep) \,\, \glog(B(\ep)) \\
        &=&  -\int d\ep \,\, \omega(\ep) \int_0^{B(\ep)} dx \,\, \glog(x)  +  \bar c\quad ,
 \end{array}
\label{newent}
\end{equation}
with $\bar c$ an integration constant which is only different from $c$, iff
$\lim_{x\to 0} x\glog(x) \neq 0$. Note immediately that the 
classical entropy is a special case of Eq. (\ref{newent}),  i.e. taking $\glog(x)=\log(x)$, 
yields the Boltzmann entropy modulo an irrelevant additive constant, 
$S[B]= - \int d\ep \,\,  \omega(\ep) \,\, B(\ep) \,\, \log(B(\ep)) + \bar c+1$. 

It can now easily be checked that this entropy Eq.(\ref{newent}),  in combination with 
the standard maximum entropy principle under the {\it usual} constraints,   
yields the measured distributions $B$. Let us define
\begin{equation}
 \begin{array}{lll}
 G = S[B] &-& \beta  \int d\ep \,\, \omega(\ep) \,\, B(\ep) \,\, (\ep-U) \\
          &-& \gamma ( \int d\ep \,\, \omega(\ep) \,\, B(\ep)-1 )  \quad , 
 \end{array}
\label{funct2}
\end{equation}
and vary with respect to $B$, to get 
\begin{equation}
\begin{array}{lll}
\frac{\delta}{\delta B}G &=&  \omega(\ep) B(E) \glog'(B(E))
                          - \frac{d}{d B} \omega(E) B(E) \glog(B(E)) \\
                        & - &\omega(E)\gamma  -\omega(E) \beta(E -U) =0 
\quad , 
\end{array}
\label{vari}
\end{equation}
or equivalently, $\glog(B(E))=-\gamma-\beta(E-U)$. Using that 
$\gexp$ is the functional inverse of $\glog$, the correct generalized
Boltzmann factor,   $B(E)=\gexp(-\beta(E-U)-  \gamma)$,   is recovered.  

%Equation (\ref{vari})  illustrates why $\eta$ has the form suggested in Eq. (\ref{ansatz}).
%If instead of $\glog$ in Eq. (\ref{gge}), one uses a functional generalization realized by,
%\begin{equation}
%\glog(B) \rightarrow \bar \glog[B]\equiv   \glog(B) + \eta[B] \quad .
%\end{equation} 
%the extra term in the entropy appears as written in Eq. (\ref{entr}), because $\eta$ is a constant 
%with respect to $\epsilon$ integration and because of the normalization condition. 
%The idea is that after  variation with respect to $B$, the additional term 
%$\frac{\delta}{\delta B} \eta[B]$, can be used to cancel the term $-\omega(E) B(E) \frac{d}{dB} \glog(B(E))$. 
%The corresponding condition, $\frac{\delta}{\delta B} \eta[B]=\omega(E) B(E) \frac{d}{dB} \glog(B(E))$, 
%immediately implies the form of $\eta$ in Eq. (\ref{ansatz}).

\section{Thermodynamics}

To show that the proposed entropy of Eq. (\ref{newent}) is fully consistent with 
the expected thermodynamic relations, 
differentiate Eq. (\ref{newent}) with respect to $U$ 
and get
\begin{equation}
 \frac{\partial}{\partial U} S[B]= \beta   \quad.
\end{equation}
Note, that the thermodynamics here is simply $dU=TdS$, since no further 
assumptions have been made on other measurements
neither in terms of thermodynamic potentials (e.g. $-PdV$ or $-\mu dN$) 
nor other (experimentally controllable) macro-state variables. 

%We do not assume to know the Hamiltonian $H$ governing  the
%observed multiplicity $\omega_1(E)$,  
%however, we remark that the Gibbs-Entropy is a constant of 
%the undisturbed dynamics described by $H$. A 
%variation of the expected energy $U$ can not be performed 
%without disturbing the dynamics, and 
%the Gibbs-Entropy behaves as one would expect 
%from a thermodynamic entropy.

Finally, if one wants to write the proposed entropy Eq. (\ref{newent}) in a form that is
suggested by the classical Gibbs form one can, by defining $L$, of course write
\begin{equation}
\begin{array}{lll}
S[B]  &= &       -  \int d\ep \,\, \omega(\ep) \int_0^{B(\ep)} dx \,\, \glog(x) \\
      & \equiv & -  \int d\ep \,\, \omega(\ep) \,\, B(\ep) L (B(\ep))
\quad ,
\end{array}
\label{rela1}
\end{equation}
which implies the relation 
\begin{equation}
L(a)= \frac1a \int_0^a dx \,\, \glog(x) \quad . 
\label{rela2}
\end{equation}
It is maybe interesting to note that $L$ is nothing but the 
mean value of the $\glog$. Of course, in general $L$ is not an 
inverse of $B$.

\section{Examples}

{\it Example: Classical  Boltzmann distributions.} 
If the experimentally measured tail of a distribution is of Boltzmann type,
$B(E)\sim \exp(-\beta E)$, then $ \glog (B) \sim \log(B)$,
and by using Eq. (\ref{rela2}), $L(B)= \frac{1}{B}( B \log( B) - B)$, 
which when put into Eq. (\ref{rela1}), yields the Boltzmann entropy, 
$S[B]= -\int d\ep \,\, \omega(\ep) \,\, B(\ep) \,\, \log(B(\ep)) +1$.

{\it Example: Asymptotic power-law distributions.}
If an experimental distribution of a $q$-exponential is observed as frequently done in 
complex systems, i.e. 
$B(E)=\left[1-(1-q) E \right]^{\frac{1}{1-q}}$. Thus the generalized logarithm is the 
so-called $q$-$\log$, 
$\glog(B) =  \log_{q}(B) \equiv  \frac{B^{1-q}-1}{1-q} $.
Inserting as before gives the Tsallis entropy \cite{tsallis88,tsallis05} times a factor,
\begin{equation}
   S[B] = - \frac{1}{2-q}  \int d\ep \,\, \omega(\ep) \,\, B(\ep) \,\, \log_q(B(\ep)) +\frac{1}{2-q}
   \quad ,
%\end{array}
\end{equation}
where we require $q<2$. The factor can in principle be absorbed into a transformation 
of $\beta$ and $\gamma$.
At this point it is also obvious that in the case of power law distributions 
the question of normalizability can become an issue.
Notice, however, that since not $B$ but $\rho=\omega B$ has to be normalizable
an implicit regularization is provided by
the maximal energy $E_{\rm max}$ that the observable system, represented by $\omega$,
can assume.

\section{Conclusion}

We started by relaxing the restriction that the Boltzmann factor 
has to be of exponential form, to allow other types of observed distributions, $B(E)$, as well. 
By doing so we introduce corresponding generalized logarithms, $\glog$ (as inverses of $B$), and suggest 
to construct the entropy of systems leading to non-exponential distributions, 
as $S=-\int d\ep \,\, \omega(\ep) \int_0^{B(\ep)} dx \,\, \glog(x)$. This is 
nothing but replacing the $p\log p$ term in the usual entropy by 
the integral, $\int \glog(p)$. Obviously classical Boltzmann-Gibbs entropy 
is obtained for the special case of $\glog(x)=\log(x)$.
We demonstrate that this entropy leads to the correct thermodynamics of the system, 
and the observed distribution functions are derived naturally from the 
maximum entropy principle with the usual constraints.
Further we show that this entropy can be written as a standard generalized 
Gibbs entropy ($\int B \glog B$) with adding a constant which is functionally 
dependent on the measured distribution \cite{lutsko}. This term somehow captures numbers 
of states in the system, which may depend on parameters like temperature. 
The functional form of measured distributions, which is a kind of knowledge about the system, 
is thus naturally fed into the definition of the entropy of the system. 

A further detail in the proposed entropy definition is that it does not contain 
any additional parameters, once the distribution is known. Once given the 
data, there is no more freedom of choice of the generalized logarithms, 
nor of the functional form of the constant.

We thank J. Lutsko for a number of important suggestions.

\bibliographystyle{unsrt}

\end{document}